\documentclass[twocolumn,PRL,showpacs,preprintnumbers,amsmath,amssymb]{revtex4-1}
\usepackage{ulem}
\usepackage{color}
\usepackage{graphicx}
\begin{document}

%
\title{Direct observation of self energy signatures of the resonant collective mode in Bi$_2$Sr$_2$CaCu$_2$O$_{8+\delta}$}
%
%
%
\author{Daixiang Mou}
\author{ Adam Kaminski}
\email{kaminski@ameslab.gov}
\affiliation{Division of Materials Science and Engineering, Ames Laboratory, Ames, Iowa 50011, USA
\\Department of Physics and Astronomy, Iowa State University, Ames, Iowa 50011, USA}

\author{Genda Gu}
\affiliation{Condensed Matter Physics and Materials Science Department, Brookhaven
National Laboratory, Upton, New York 11973, USA}

%

\begin{abstract}
We use high resolution ARPES to study the resonant, collective excitation mode in the superconducting state of Bi2212. By collecting very high quality data we found new features in the self energy in the antinodal region, where the interaction of electrons with the mode is the strongest. This interaction leads to pronounced peak in the scattering rate and we demonstrate that this feature is directly responsible for well known peak-dip-hump structure in cuprates. By studying how the weight of this peak changes with temperature we unequivocally demonstrate that interaction of electrons with resonant mode in cuprates vanishes at T$_c$ and is very much localized in the momentum space close to the antinode. These findings present a consistent picture of line shape and self energy signatures of the electron-boson coupling in cuprates and resolve long standing controversy surrounding this issue. The momentum dependence of the strength of electron-mode interaction enables development of quantitative theory of this phenomenon in cuprates.
\end{abstract}
\maketitle

\section{INTRODUCTION}

In classical superconductors, the pairing and superconductivity is caused by interaction of electrons with phonons \cite{Maxwell1950,Reynolds1950,Bardeen1957}. In unconventional superconductors such as cuprates \cite{Damascelli2003, Campuzano2004} or pnictides \cite{Johnston2010,Canfield2010}, the origin of pairing is still debated, and a significant effort is made to identify the boson (if one exists) responsible for pairing. Several electron-boson interactions were discovered by ARPES in cuprates  \cite{Shen1997, Norman1997, Campuzano1999, Bogdanov2000, Kaminski2001, Johnson2001, Lanzara2001, Zhou2003, Sato2003, Borisenko2003, Cuk2004, Kordyuk2006, Lee2009, Kondo2013}, however their role in the mechanism of the high temperature superconductivity remains unknown. The strongest of these interactions causes clearly visible features in the spectral line shape leading to famous ``peak-dip-hump" structure \cite{Shen1997, Norman1997, Norman1998, Campuzano1999, Zeyher2001, Eschrig2003, Eschrig2006} and features in Raman spectra\cite{Zeyher2013, Sacuto2017}. This mode is attributed frequently to  so called resonant mode first observed by inelastic neutron scattering \cite{Rossat-Mignod1991,Mook1993, Fong1999}, although some argued that this is a signature of very strong electron-phonon coupling \cite{Lanzara2001, Gweon2004, Cuk2004, Devereaux2004}. Some of the early papers advocating the phonon origin of this effect were based on seemingly isotropic behavior around the Fermi surface \cite{Lanzara2001} and observation of isotope effect \cite{Gweon2004} at high binding energies. Other result clearly demonstrated that this interaction varies strongly with momentum and temperature and is strongest in the antinodal region, where the superconducting gap reaches maximum magnitude \cite{Kaminski2001, Cuk2004, Devereaux2004}, although there were some arguments that phonons can also couple to electrons in highly anisotropic manner \cite{Cuk2004, Devereaux2004}. The isotope effects studies so far still lack consensus \cite{Gweon2004, Douglas2007,Iwasawa2007,Iwasawa2008}. A theoretical model also proposed that electron-phonon interaction can indeed change abruptly across T$_c$ \cite{Devereaux2004}, although this does not occur in classical superconductors \cite{Mou2015}.

\begin{figure}[htbp]
\centering
\includegraphics[width=\columnwidth]{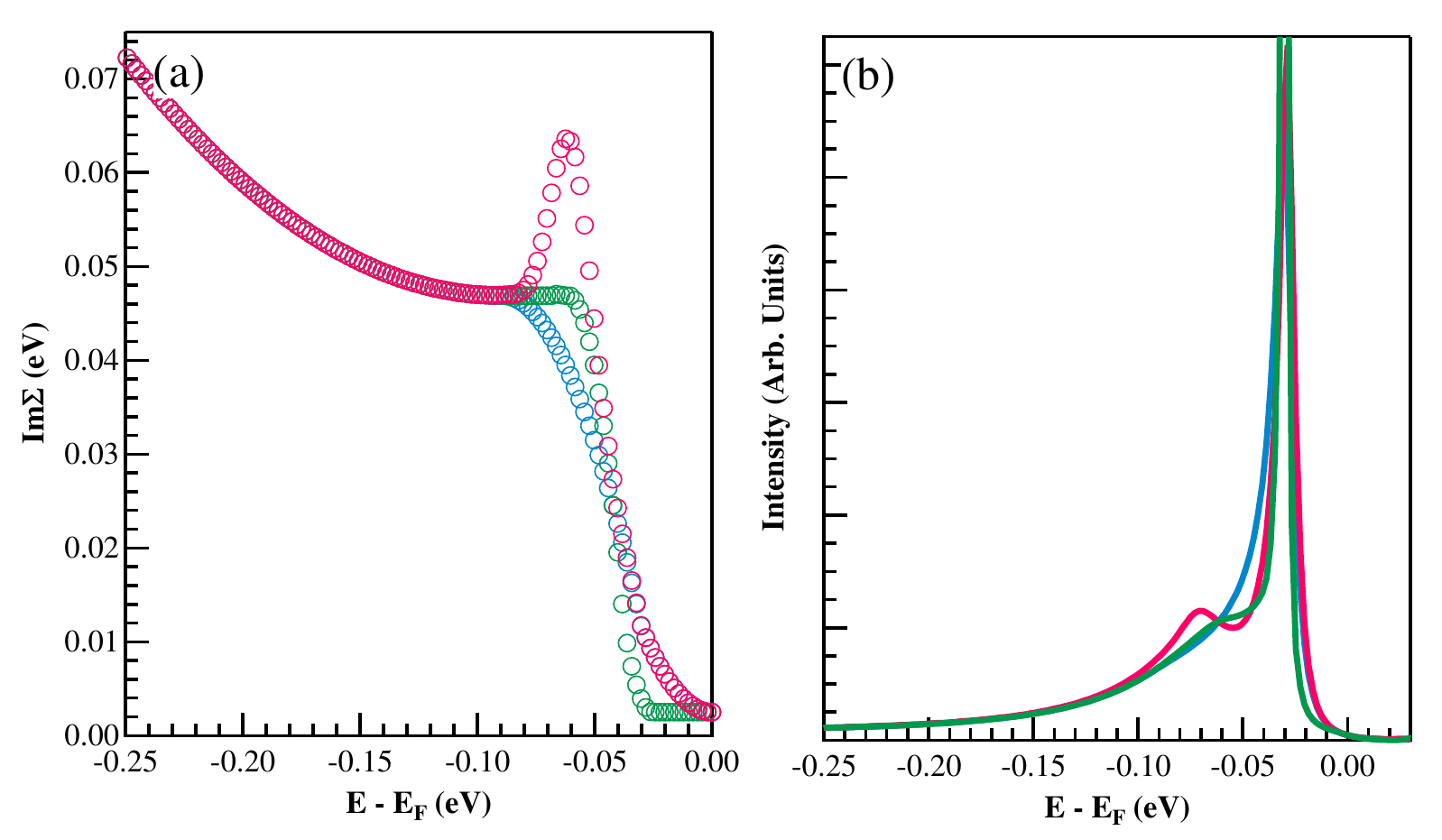}
\caption{(Color) EDC line shape simulation. (a) three different Im$\Sigma$ models, two with a broadened step-like functions (blue and green circles) and one with a peak structure (red circle) at 60meV. (b) Simulated k$_F$ EDCs in superconducting state. Gap size of 30 meV is used in all simulations. Detailed simulation procedure is described in Ref. \cite{Mou2015}. }
\end{figure}

Presence of a collective excitation in a material has pronounced effects on both real (Re$\Sigma$) and imaginary (Im$\Sigma$) part of the self energy \cite{Valla1999, Kaminski2001}, a quantity that can be readily extracted from ARPES data \cite{Valla1999}. Indeed, a significant progress was made in the nodal region of the momentum space using such approach \cite{Shen1997, Norman1997, Bogdanov2000, Kaminski2001, Johnson2001, Lanzara2001, Zhou2003, Sato2003, Borisenko2003, Cuk2004, Kordyuk2006, Lee2009, Kondo2013}.  The main obstacle in examining the detailed temperature dependence at the antinode is lack of objective measure of its strength due to rapid broadening of the peaks above T$_c$ and the proximity of the band bottom to the chemical potential at antinode. Here we solve this decades, longstanding problems by using double Lorentzian fits to ultra high quality and statistics data. This approach allowed us to successfully extract the self energy in the antinodal part of the Brillouin zone and resolve this important issue once and for all. We are not aware of other previous attempts to extract the Im$\Sigma$ at the antinode. It is possible that requirement of very high data statistics was a barrier for experiments carried out at synchrotron due to time limits and sample aging \cite{Palczewski2010}. We found very pronounced peak in the Im$\Sigma$ in the close proximity of the antinode at energy expected based on the inelastic neutron scattering studies. Upon warming up, the amplitude of this peak decreases in order parameter-like fashion and it disappears completely above T$_c$. The temperature dependence of this feature follows exactly the intensity of the resonant mode reported by neutron measurements. The momentum dependence of this feature will provide new insights and guidance to theory to establish the relation of this effect to the high temperature superconductivity in cuprates.

\begin{figure}[htbp]
\centering
\includegraphics[width=0.9\columnwidth]{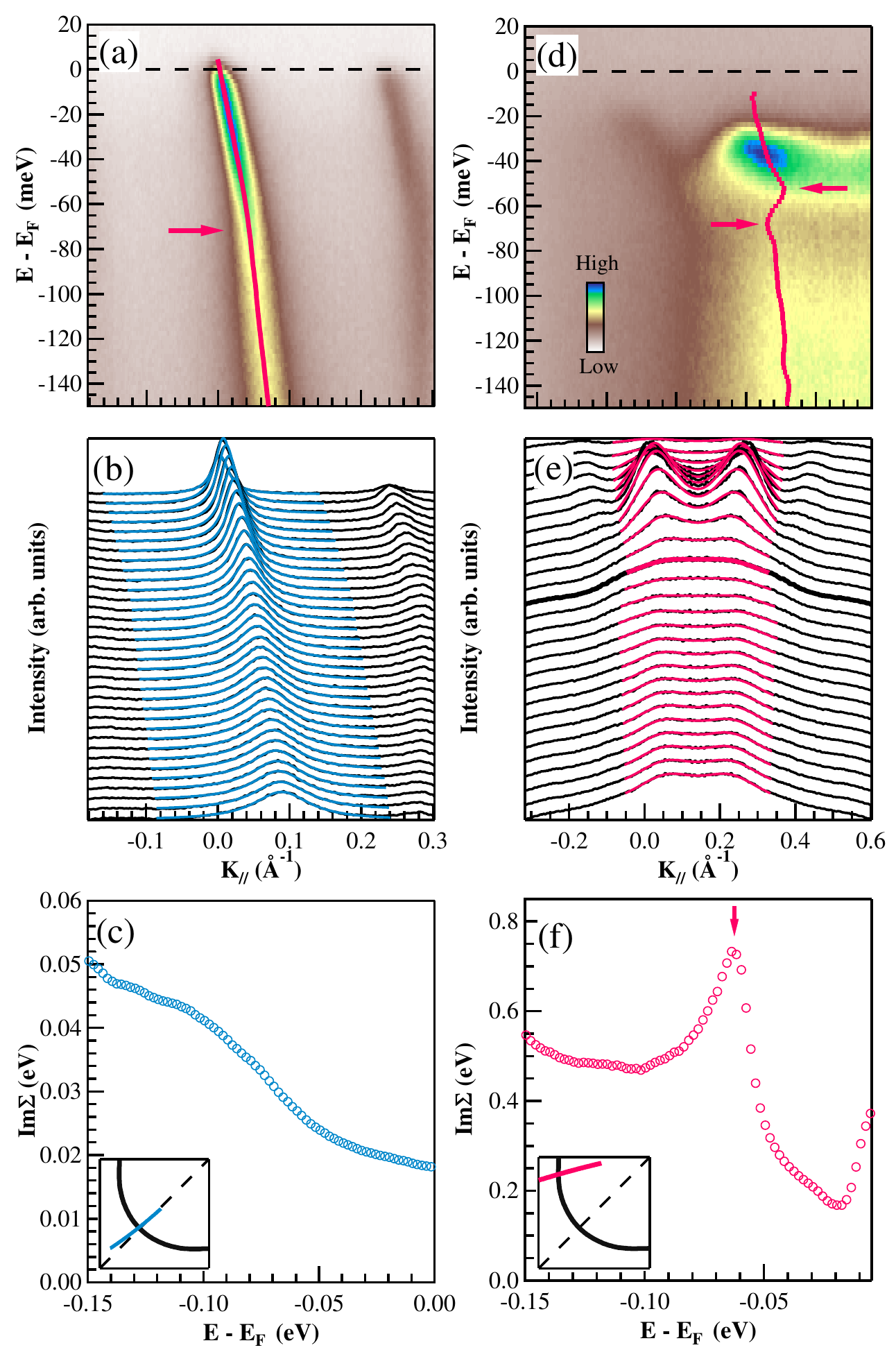}
\caption{(color)  Typical MDC fitting results. (a) Measured spectra image along nodal cut at 16 K. Cut location is illustrated in the insert of (c). Obtained band dispersion is superimposed on the image (blue solid line). (b) Corresponding MDCs (black solid lines) and fitting results (blue solid lines). Data are shifted vertically for clarity. (c) Energy dependent Im$\Sigma$ of the cut in (a). (d) - (f) Similar as (a) - (c) with cut near antinodal region, as illustrated in the insert of (f).  The MDCs in (e) are symmetrized from left to right with the momentum of band bottom . Thick black line marks the MDC at the energy of peak position in (f). }
\end{figure}

\section{EXPERIMENTAL DETAILS}

Optimally doped single crystals of Bi2212 (T$_c$ = 95 K) were grown by the conventional floating-zone technique and used in number of previous studies \cite{Kondo2009, Kondo2011, Kondo2013b}. We note that the T* for these samples is around  230K, therefore substantially higher than T$_c$ as demonstrated in Ref. \cite{Kondo2013b}.  The samples were cleaved {\it in situ} at base pressure lower than 5 $\times$ {10$^{-11}$}  Torr. The cleaved surface displayed no observable aging effects for duration of measurements as verified by temperature cycling. ARPES measurements were carried out using a laboratory-based system consisting of a Scienta SES2002 electron analyzer and GammaData Helium UV lamp equipped with custom designed refocusing optics. All data were acquired using the HeI line with a photon energy of 21.2 eV. The angular resolution was 0.13$^\circ$ and $\sim$ 0.5$^\circ$ along and perpendicular to the direction of the analyzer slits, respectively. The energy corresponding to the chemical potential was determined from the Fermi edge of a polycrystalline Au reference in electrical contact with the sample. The energy resolution was set at $\sim$6meV - confirmed by measuring the energy width between 90\% and 10\% of the Fermi edge from the same Au reference. The data were measured using several samples yielding consistent results.

\begin{figure}[htbp]
\centering
\includegraphics[width=\columnwidth]{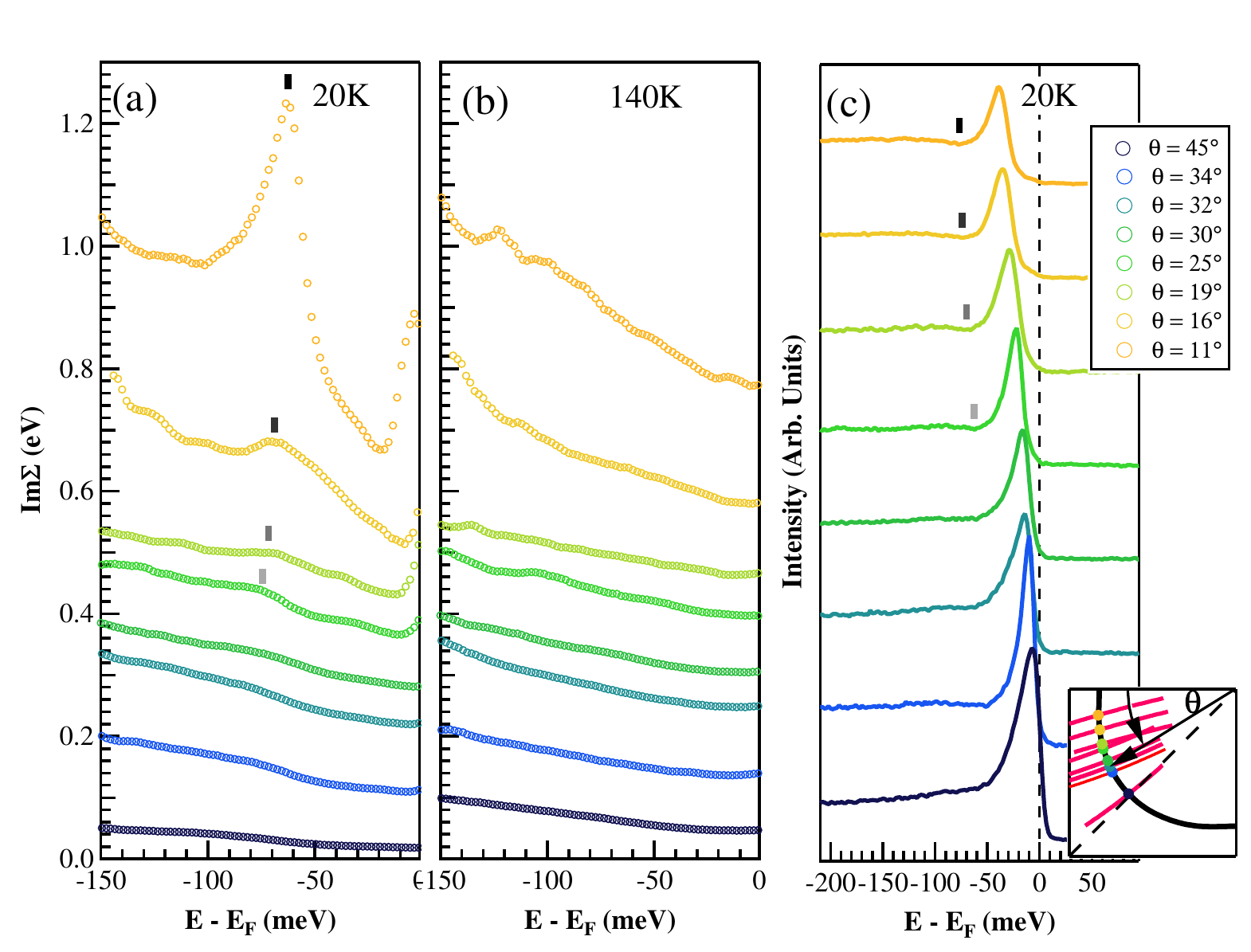}
\caption{(Color) Extracted energy dependent Im$\Sigma$ along different cuts in Brillouin Zone at  superconducting state 20 K (a) and normal state 140 K (b). Cut locations and $\theta$ definition are illustrated in right insert. Data are offset vertically with an 150 meV interval for each cut. (c) Extracted k$_F$ EDCs from each cut measured at 20 K. Data are offset vertically for clarity. The black vertical bars in (a) and (c) mark the peak positions of Im$\Sigma$  and the dip positions in EDC.}
\end{figure}

\begin{figure}[htbp]
\centering
\includegraphics[width=0.9\columnwidth]{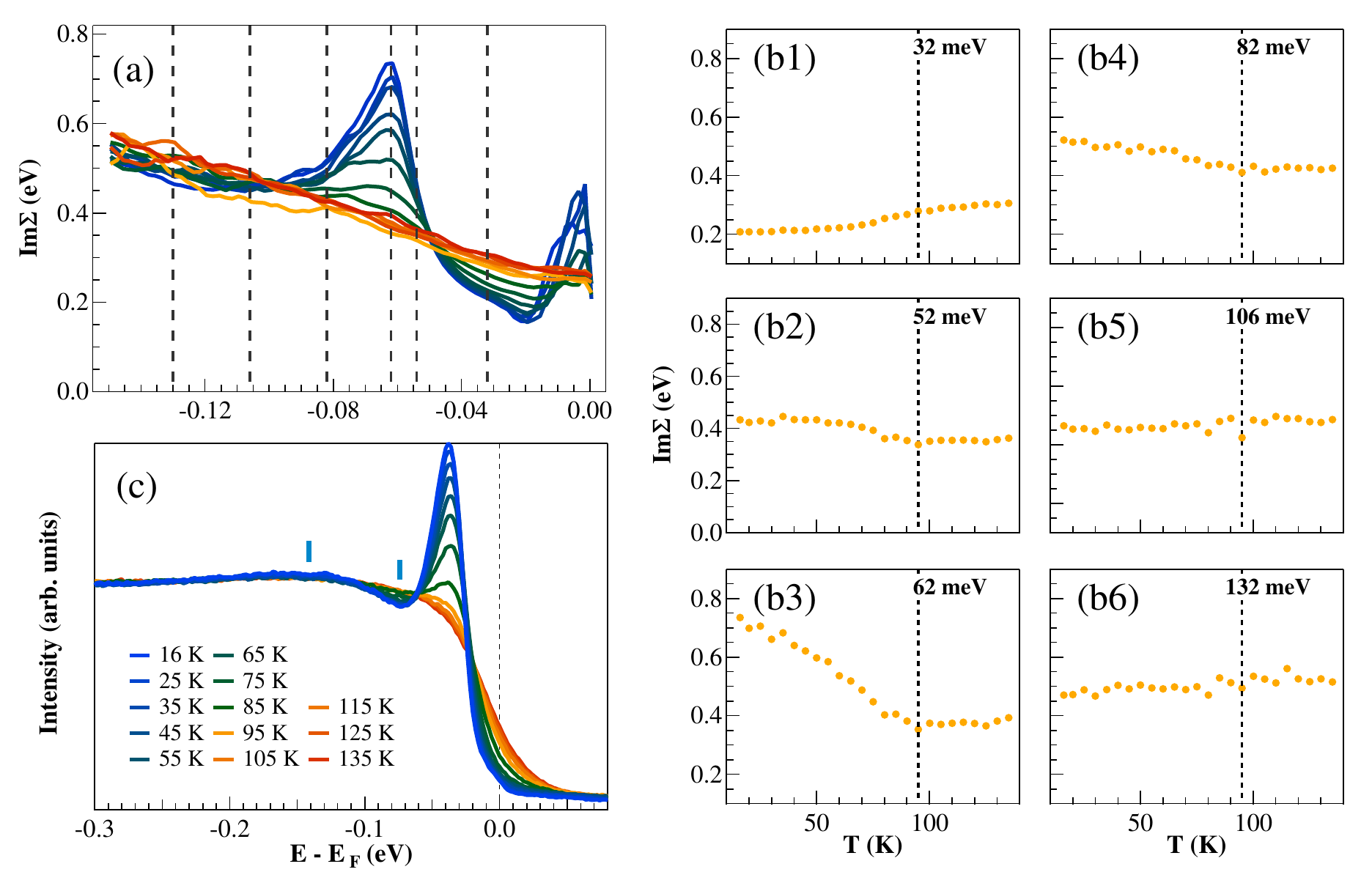}
\caption{(Color) Temperature dependent Im$\Sigma$ of $\theta=11^{\circ}$ cut. (a) Energy dependent Im$\Sigma$ at different temperatures.  (b) Temperature dependent Im$\Sigma$ at binding energy of 32 meV, 62 meV, 92 meV, 106 meV and 132 meV. {Vertical doted line marks T$_C$.} (c) k$_F$ EDCs at different temperatures. The dip and hump positions are marked. }
\end{figure}

Im$\Sigma$ is obtained by fitting the Momentum Distribution Curve (MDC) with Lorenzians \cite{Valla1999, Kaminski2001}. The half width of Lorenz peak ($\Delta$k) is proportional to Im$\Sigma$: $Im\Sigma = \Delta k  \times \nu_0$, where $\nu_0$ is the bare Fermi velocity. At the node this is a straight forward procedure as there is only a single, relatively sharp, highly dispersive band present. Close to the antinode such procedure becomes much more complicated because the bottom of the band is located close to the chemical potential and relatively broad peaks for positive and negative momentum partially overlap. To overcome these problems, we fitted the MDC data at each binding energy with two lorentzians. We took care verifying that the dispersion determined by the positions of these peaks agrees with the overall data. Such obtained dispersion data allows us also to determine the velocity that is necessary for calculating the Im$\Sigma$.

The presence of the peak in the Im$\Sigma$ is a necessary condition for the appearance of the peak-dip-hump structure that is a hallmark of the ARPES line shape in the superconducting state of cuprates. We demonstrate this in Fig. 1a, where we model an Im$\Sigma$ that has an abrupt suppression (two examples with different slope)  below the mode energy and one that has an additional peak at the same energy. We than calculate the EDC intensities using standard expression for the spectral function and Re$\Sigma$ obtained by Kramers Kronig transformation of the assumed Im$\Sigma$'s. The results are shown in panel (b). In the first two cases, the resulting EDC has only a shoulder at the energy of the mode and no decrease of the intensity occurs. In the third case  however, the presence of the peak in the Im$\Sigma$ causes a clear dip, in qualitative agreement with hallmark EDC ARPES data lineshape. We therefore demonstrate that the presence of the dip in the EDC implies that the { Im$\Sigma$} must have a peak at similar energy.

\begin{figure*}[htbp]
\centering
\includegraphics[width=0.8\textwidth]{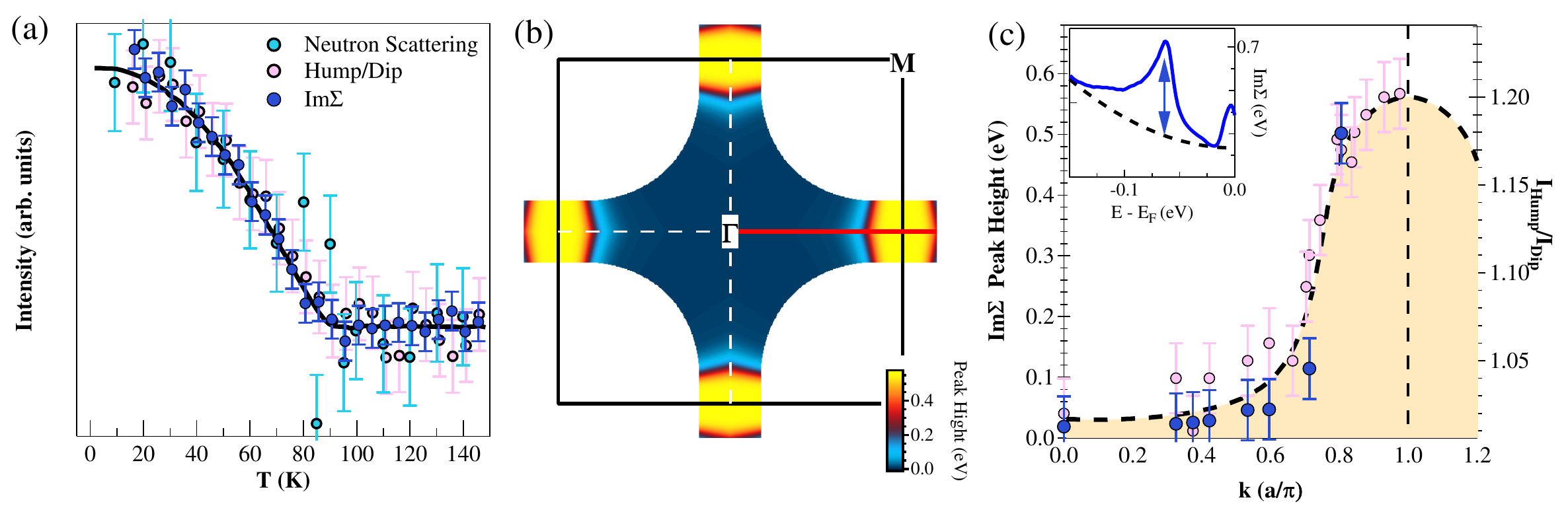}
\caption{(Color) (a) Temperature dependence of resonance mode intensity from neutron scattering (cyan solid circles)(Ref. \cite{Fong1999}), Im$\Sigma$ at 62 meV ( blue solid circles) and dip to hump intensity ratio in k$_F$ EDCs obtained from data in fig. 4c ({pink} solid circles). Data are shifted and normalized for comparison. The black line is a guide to the eye. (b) Momentum dependent peak {height} in Im$\Sigma$. (c) Peak { height} in Im$\Sigma$ along $\Gamma$-X cut. As illustrated in the insert, peak hight height is obtained by subtracting a fitted Im$\Sigma$ with formula $a\omega^2+b$ (black dashed line in insert). }
\end{figure*}

\section{RESULTS AND DISCUSSION}

We now proceed to experimentally extract this information from our ultra high quality data. The key results along the node and antinode for T=16 K, well below T$_c$ are shown in Fig. 2. In panels a, b and c we show the ARPES data along the node, fits to the MDCs and extracted Im$\Sigma$ respectively. The nodal data and analysis are very similar to previous results demonstrating high degree of consistency. We observe several small bumps at high and low binding energies. Those most likely represent very weak coupling to ever-present phonons. Below, in panels d-f we present equivalent analysis of the data measured close to the antinode. The antinodal data is far more interesting, as such analysis was not carried out previously. The MDC's at high binding energy (bottom of panel e) are relatively broad, however it is clear that they consists of two separate peaks. At lower binding energies, the MDC marked as a bold line looks much more like a single broader peak and one can no longer easily distinguish two separate peaks. This is a clear signature that the line width of the two peaks that originate from the two branches for positive and negative momentum become broader than their separation at this binding energy and the overall MDC looks more like a single, {broad} peak. This agrees with the detailed results of extracted Im$\Sigma$ shown in panel f. The Im$\Sigma$ initially decreases with decreasing binding energy, then increases to nearly double of its initial value, reaching a peak at binding energy of 62 meV. At even lower energies, approaching the E$_F$, the Im$\Sigma$ decrease rapidly to value well below one at high binding energies as expected. The upturn of the Im$\Sigma$ near E$_F$ is caused by superconducting gap opening.


We now move to study how this feature in Im$\Sigma$ evolves with momentum by carrying out similar measurements and analysis along the entire Fermi surface. The results are summarized in Fig.~3. In the superconducting state (Fig.~3a), the previously mentioned peak structure can be clearly observed near antinode and is quite rapidly suppressed away from this part of the Brillouin zone with Im$\Sigma$ evolving into a step-like structure around $\theta=25^{\circ}$ that remains to and is typical for the nodal direction.
The energy position of the peak is shifting towards the higher binding energy as we move away from the antinode, indicating smooth variation of the mode energy coupled to electrons along Fermi surface. This evolution is consistent with the recent studies of the Re$\Sigma$ using laser ARPES  away from the antinodal region.\cite{He2013,Plumb2013}.
To contrast the behavior in the superconducting state, we show the extracted Im$\Sigma$ along the same cuts in the normal state (140 K) in Fig.~3b. No obvious energy scale can be identified for any of these cuts signifying that these effects are somehow related to the superconducting state.

We now proceed to study in detail the temperature dependence of the self energy in the antinodal region. The extracted Im$\Sigma$ for $\theta=11^{\circ}$ cut plotted as a function of temperature are shown in Fig.~4a. The peak in Im$\Sigma$ is gradually suppressed with increasing temperature and vanishes above T$_c$. We extract Im$\Sigma$ at selected binding energies and plot them as a function of temperature in Fig.~4b.
It is clear from these plots that the temperature dependence of the Im$\Sigma$ changes qualitatively with binding energy. At low banding energies of 32 meV (slightly below the superconducting gap energy), the Im$\Sigma$ is suppressed below T$_c$. This is consistent with earlier studies of the nodal region as one would expect equivalency between temperature and binding energy in the self energy.
In contrast, at binding energy of 62 meV (peak position in Im$\Sigma$ - see Fig.~4a), Im$\Sigma$ begins to increase upon cooling  just below T$_c$ and then follows a BCS-like temperature dependence (panel b3). The increase of Im$\Sigma$ upon cooling vanishes for higher binding energies as shown in panel b5 and b6 and the Im$\Sigma$ does not display significant changes with temperature as expected.

We summarize the temperature and momentum dependence of the peak structure in Im$\Sigma$ in Fig. 5. The comparison between Im$\Sigma$ for binding energy of 62 meV and  the ratio of the intensity at dip and hump energies are shown in Fig.~5a. Clearly they have the same temperature dependence that closely follows the intensity of the resonant peak measured in the inelastic neutron scattering experiments\cite{Fong1999}.  We stress that the T*$\sim$230 K temperature for those samples is significantly higher than T$_c$\cite{Kondo2013b}, therefore the signatures of the resonant mode exist only in the superconducting state. In Fig. 5b we present a color plot of the peak value of the Im$\Sigma$ throughout the Fermi surface. Fig. 5c shows how this quantity evolves along the $\Gamma$-{X} symmetry direction. The very high values of the Im$\Sigma$ observed only in close proximity to the antinode, which is very consistent with the formation of the spin resonance mode in this  region of momenta\cite{Fong1999}.

The discovery of the peak structure in Im$\Sigma$, together with its momentum and temperature dependence,  provides a natural explanation of the peculiar spectra line shape observed in Bi2212 for more than two decades. Above T$_c$, in the antinodal region, EDC show a very broad peak, which is a hallmark of the normal state of optimally and underdoped cuprates. Upon cooling below T$_c$, a sharp quasiparticle peak forms at the energy corresponding to the superconducting gap and is followed by a dip then a hump at higher binding energies (Fig.~4c) \cite{Dessau1991}. The key proposal explaining the dip and hump structure relies on the interaction of electrons with a collective excitation \cite{Dessau1991, Shen1997,Norman1997, Norman1998}. Such proposal necessarily requires existence of peak in the Im$\Sigma$. Our data provides definite evidence of its existence and the one to one correspondence between the peak  in Im$\Sigma$ and the dip-hump structure in EDC line shape validates those theory proposals. Our data reveals highly anisotropic momentum dependence of the enhancement of the Im$\Sigma$. This feature is almost completely extinguished within 20\% of the M-$\Gamma$ distance.

In summary, we describe a systematic investigation of the self energy in the antinodal region of optimally doped Bi2212. We find direct evidence of strong enhancement of the scattering rate in the superconducting state that manifests itself as a sharp peak in the Im$\Sigma$. We demonstrate that this feature is directly responsible for { peak-dip-hump} structure that attracted a lot of interest due to its intimate relation with superconductivity. The energy, temperature and momentum dependence of this feature matches very well with properties of the resonant mode reported by inelastic neutron scattering experiments. The observed peak in Im$\Sigma$ vanishes around T=$\sim$90K, which is similar to T$_c$ for these samples and much lower than T* (=230K). This demonstrates that the resonant mode is linked with T$_c$ rather than T*. The quantitative information about the momentum dependence of the strength of coupling to this collective excitation will provide guidance and validation of the theories explaining its detailed origin and relation to high temperature superconductivity.

This work was supported by the US Department of Energy, Office of Basic Energy Sciences, Division of Materials Sciences and Engineering. Ames Laboratory is operated for the US Department of Energy by the Iowa State University under Contract No. DE-AC02-07CH11358. 
\bibliography{Bi2212_SelfEnergy}

\end{document}